\documentclass[aps,prl,reprint,groupedaddress]{revtex4-1}
\usepackage {graphicx}  

\usepackage{epstopdf}
\usepackage{bm}        
\usepackage{amssymb}   

\makeatletter

\newcommand{\Rmnum}[1]{\expandafter\@slowromancap\romannumeral #1@}
\makeatother

\begin{document}

\title{The $\nu=5/2$ Fractional Quantum Hall State in the Presence of Alloy Disorder}

\author{Nianpei Deng$^1$, G.C. Gardner$^{2,3}$, S. Mondal$^{1,2}$, E. Kleinbaum$^{^1}$, M.J. Manfra$^{1,2,3,4}$, and G.A. Cs\'{a}thy$^{1,2}$ \footnote{gcsathy@purdue.edu}}

\affiliation{${}^1$ Department of Physics, Purdue University, West Lafayette, IN 47907, USA \\
${}^2$ Birck Nanotechnology Center, Purdue University, West Lafayette, IN 47907, USA  \\
${}^3$ School of Materials Engineering, Purdue University, West Lafayette, IN 47907, USA \\
${}^4$ School of Electrical and Computer Engineering, Purdue University, West Lafayette, IN 47907, USA \\}

\date{\today}

\begin{abstract}
We report quantitative measurements of the impact of alloy disorder on the $\nu=5/2$ fractional quantum Hall state.  
Alloy disorder is controlled by the aluminum content $x$ in the Al$_x$Ga$_{1-x}$As channel of a quantum well.
We find that the $\nu=5/2$ state is suppressed with alloy scattering. 
To our surprise, in samples with alloy disorder the $\nu=5/2$ state appears at significantly reduced mobilities
when compared to samples in which alloy disorder is not the dominant scattering mechanism.
Our results highlight
the distinct roles of the different types of disorder present in these samples, such as the
short-range alloy and the long-range Coulomb disorder.

\end{abstract}

\pacs{}
\maketitle

Unraveling the impact of disorder is an important endeavor in contemporary condensed matter physics. 
Disorder is well understood in the single particle regime.
Examples of fundamental importance are Anderson localization \cite{Anderson} 
and universal plateau-to-plateau transition in the integer quantum Hall effect \cite{Wanli05}.
Localization in the presence of the disorder is also important
in topological insulators \cite{Andrew} and in atomic condensates \cite{BEC}. 
In contrast, understanding disorder in correlated electron systems continues to pose serious challenges.
The interplay of disorder and interactions has witnessed renewed interest 
in the two-dimensional electron gas (2DEG) in connection to the stability
of the exotic fractional quantum Hall states (FQHS) \cite{Umansky,Pan11,Gamez}
and in graphene due to the observation of a wealth of FQHSs using local detection \cite{Graphene}.

The FQHS at Landau level filling factor $\nu=5/2$ is one example of a correlated ground state which
has attracted considerable attention
\cite{Willett,Pan99,Xia,Eisen02,Pan08,Choi,Nuebler,Kumar,Shayegan,Dean,Nodar,Du,Pinczuk,Radu,Dolev,Bid,Willett10,Vivek}. 
This is because of the putative exotic Pfaffian-like correlation in the
ground state at $\nu=5/2$ and of the non-Abelian quasiparticle excitations \cite{Moore,Martin,Bonderson}. 
Non-Abelian quasiparticles
may be used to realize topological qubits, building blocks of fault-tolerant quantum computers \cite{DasSarma}. 
Furthermore, since the Pfaffian can be mapped into a paired wavefunction with a $p$-wave symmetry \cite{Moore,Martin,Scarola},
the $\nu=5/2$ FQHS is intimately connected to $p$-wave superconductors \cite{MacKenzie}, Majorana physics in 
superconductor-semiconductor hybrid devices \cite{Sau}, and superfluid $^3$He \cite{Leggett}.

The effect of the disorder on the $\nu=5/2$ FQHS remains largely unknown \cite{Umansky,Pan11,Gamez}.
Disorder is a key factor in limiting $\Delta_{5/2}$, the energy gap of the $\nu=5/2$ FQHS, to less than 0.6~K
\cite{Pan08,Choi,Kumar}. Measurements of this state must therefore be conducted at
either dilution or nuclear demagnetization refrigerator temperatures, which render these
studies time consuming \cite{Pan99,Zumbuhl}.
However, in the disorder-free limit $\Delta_{5/2}$ is predicted to be as high as 2~K
at the typical electron density of $n=3 \times 10^{11}/$cm$^2$ \cite{Morf02,Morf,Wojs,Peterson,Feiguin}.
Understanding disorder in the $\nu=5/2$ FQHS is thus expected to lead to an increased 
energy gap with the following benefits toward fundamental tests of the nature of this state:
a) experiments may be conducted at higher temperatures, possibly in $^3$He refrigerators, 
with shorter turn-around times allowing for extensive investigations
b) improved signal-to-noise ratio in experiments on nanostructures in which the edge states of
the $\nu=5/2$ FQHS are probed \cite{Dolev,Radu,Bid,Willett10}
and c) exponentially enhanced topological protection in qubits \cite{DasSarma}.

Studies of disorder require the capability of its control. 
In this Letter we report on a quantitative inquiry of the impact of a specific type of short-ranged disorder, alloy disorder, on the $\nu=5/2$ FQHS.
We investigated a series of specially engineered samples in which all parameters but the alloy disorder
remain constant by design \cite{Gardner}. Specifically, we measured
Al$_{0.24}$Ga$_{0.76}$As/Al$_x$Ga$_{1-x}$As/Al$_{0.24}$Ga$_{0.76}$As quantum wells 
in which the electrons are confined to the Al$_x$Ga$_{1-x}$As alloy 
and which have different values of the aluminum molar fraction $x$ \cite{Gardner}. 
Since the disorder is added to the electron channel during 
the Molecular Beam Epitaxy (MBE) growth, it is controlled and precisely quantified.
Disorder is found to suppress the energy gap of the $\nu=5/2$ FQHS.
However, to our surprise we find strong $\nu=5/2$ FQHSs in alloy samples at 
values of the electron mobility
at which this state does not develop at all in the highest quality alloy-free samples.
The mobility threshold for the formation of the $\nu=5/2$ FQHS 
in the alloy samples is thus much reduced as compared to that in the alloy-free samples.
Our results indicate that the engineering of the
exotic FQHSs, such as the one at $\nu=5/2$, is critically dependent on the different length scales of 
competing disorders present in the 2DEG: the short-range alloy and interface roughness disorder
and the long-range Coulomb disorder.

A sketch of the active region of our alloy-free reference sample
and of a sample containing alloy disorder are shown in the insets of Fig.1. 
The sample growth procedure and characterization of our samples at $T=300$~mK 
can be found in Ref.\cite{Gardner}. Details of the samples, of the
state preparation, and of other
aspects of the low temperature measurements reported in this Letter can be found in the Supplementary Materials section.
We note that MBE-controlled alloy disorder was first introduced to 2DEGs in Ref.\cite{Wanli03}.
However, in these samples \cite{Wanli05,Wanli03,Wanli10} the $\nu=5/2$ FQHS 
has not been observed. In contrast, our samples have several essential features which are optimized for a 
strong $\nu=5/2$ FQHS even in the presence of the disorder.
First, the 2DEG is confined to a symmetrically doped quantum well rather than a single heterointerface. This allows
for a higher electron density, enhancing therefore fragile FQHSs.
Second, we use a reduced Al content $0.24$ in the Al$_{0.24}$Ga$_{0.76}$As barriers,
which enhances the $\nu=5/2$ FQHS \cite{Pfeiffer,Eisen02,Gamez}. Third, we use a short period superlattice doping scheme \cite{Friedland},
which is known to yield a strong $\nu=5/2$ FQHS \cite{Pfeiffer,Umansky,Manfra}. 

It is important to appreciate that only the alloy disorder is different in each sample.
All other sample parameters, however, are left virtually unchanged. In order to avoid any density dependent effects
the electron density is kept constant, close to $n \simeq 2.9\times10^{11}$/cm$^2$.
Specifically, in our samples $2.70\times10^{11}$/cm$^2 \leq    n \leq 3.08 \times10^{11}$/cm$^2$.
Furthermore, the alloy content $x$
of the electron channel Al$_x$Ga$_{1-x}$As is low when compared to that in the confining Al$_{0.24}$Ga$_{0.76}$As. 
There is therefore  virtually no variation of the electronic effective mass $m$ and of
the electronic confinement in the direction perpendicular to the plane of the 2DEG \cite{nextnano}. 
Other parameters held constant include the position of the 2DEG relative to the sample surface and the 
thickness of the capping layer \cite{laroche}. 

Fig.1 shows the magnetoresistance $R_{xx}$ and the Hall resistance $R_{xy}$ of the alloy-free reference sample,
i.e. for which $x=0$, measured at $T=7$~mK in a van der Pauw geometry.    
The figure is limited to magnetic fields $B$ for which the filling factor ranges between $2<\nu<3$, 
commonly referred to as the lower spin branch of the second Landau level.
Because of the high quality growth and sample design described earlier 
we observe strong FQHSs at $\nu=5/2$, $2+1/3$, and $2+2/3$ as indicated by vanishing $R_{xx}$ and quantized $R_{xy}$ \cite{Tsui,Willett}.
Other more fragile FQHSs are also seen in Fig.1 \cite{Xia,Kumar}. 

\begin{figure}[t]
 \includegraphics[width=0.99\columnwidth]{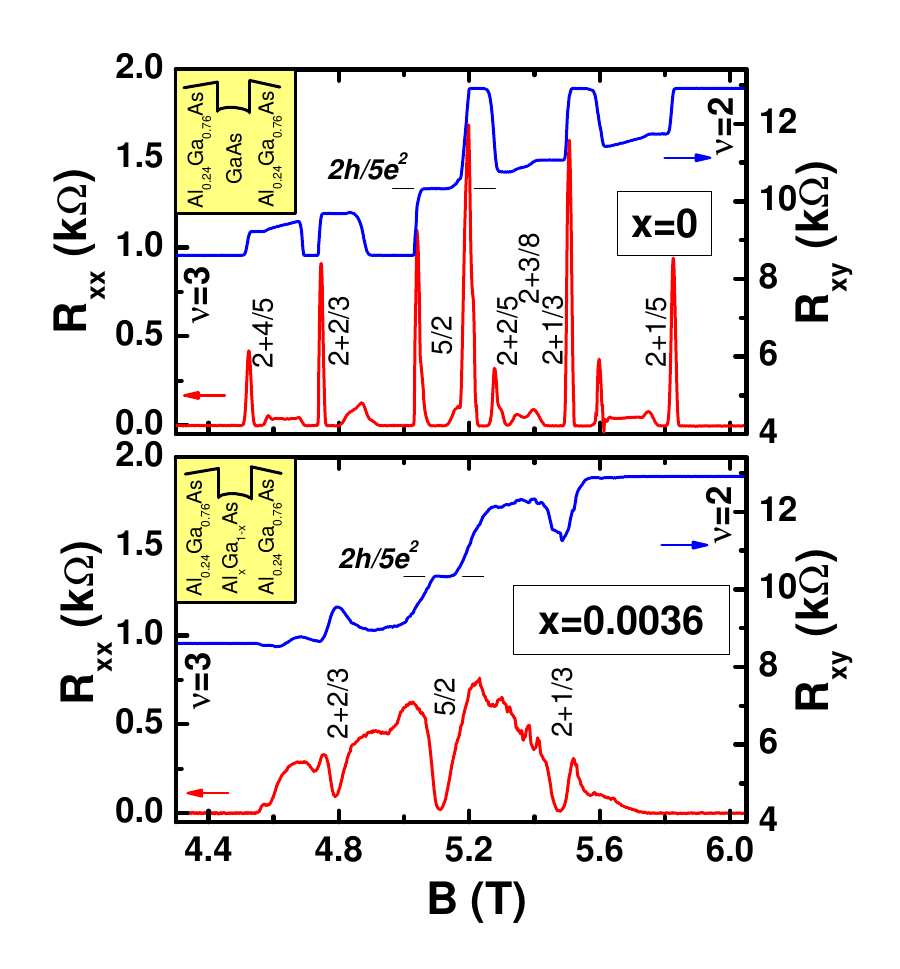}
 \caption{ Magnetoresistance $R_{xx}$ and Hall resistance $R_{xy}$ in the reference sample with $x=0$ 
 (top panel) and the alloy sample with $x=0.0036$ (bottom panel) as measured at 7~mK. Numbers indicate 
 the filling factors of various FQHSs and insets are sketches of the sample structure.
 \label{Fig1}}
 \end{figure}
 
Alloy disorder strongly affects magnetoresistance. 
This can be seen in the traces of the sample with $x=0.0036$ gathered at 7~mK, which are also shown in Fig.1. 
The most fragile FQHSs, such as the $\nu=2+2/5$, $2+1/5$, and $2+4/5$ FQHSs, are destroyed.
The FQHS at $\nu=5/2$, however, remains fully quantized
in spite of the presence of alloy disorder. Indeed, at this $\nu$ there is 
a vanishing $R_{xx}$ and the rigorously quantized $R_{xy}=2h/5e^2$, which holds to a precision of 1 part in 10$^3$.

Next we have investigated the temperature dependence of the  $\nu=5/2$ FQHS.
Thermalization of electrons in our experiment is
assured by the use of a $^3$He immersion cell \cite{Pan99,Nodar11} and temperature measurements of the $^3$He bath 
below 100~mK are performed with the aid of a tuning fork viscosity thermometer \cite{Nodar11}.
At the lowest temperatures $T$
the magnetoresistance of the $\nu=5/2$ FQHS follows an activated form $R_{xx} \propto \exp(-\Delta_{5/2}/2T)$,
from which we extract the energy gap $\Delta_{5/2}$.  
The inset of Fig.2 shows the temperature dependence of $R_{xx}$ at $\nu=5/2$ on an Arrhenius plot, 
i.e. $\ln R_{xx}$ as function of $1/T$. 
The presence of the linear segment indicates that transport is activated.
In the alloy-free reference sample we find a record high energy gap 
$\Delta_{5/2}=569$~mK \cite{Kumar}.

\begin{figure}[t]
 \includegraphics[width=0.9\columnwidth]{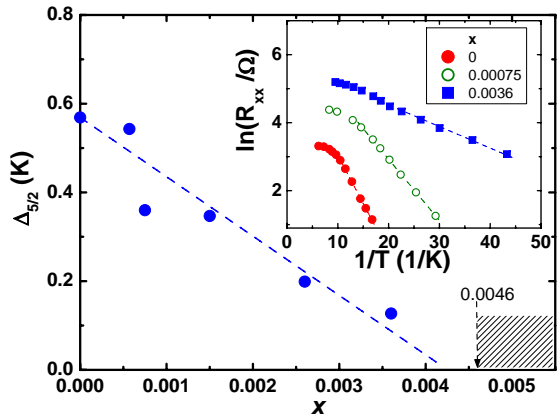} 
 \caption{ The dependence of $\Delta_{5/2}$ on the aluminum mole fraction $x$. 
  In a sample with $x=0.0046$ we do not observe a FQHS at $\nu=5/2$, hence the excluded shaded area.
 The inset shows Arrhenius plots for the $\nu=5/2$ FQHS in three representative samples.
 \label{Fig2}}
 \end{figure}
 
The inset of Fig.2 also shows the $T$-dependence of $R_{xx}$ at $\nu=5/2$
in two representative alloy samples with $x=0.00075$ and $x=0.0036$.
The presence of linear segments at non-zero $x$ means the survival
of activated transport even in the presence of alloy disorder. It is, therefore, meaningful to extract
energy gaps in the alloy samples as well. Values found are tabulated in the Supplementary Materials section
and are plotted as function of $x$ in Fig.2.
We find the $\Delta_{5/2}$ has a decreasing trend with an increasing $x$.
At the largest value of  $x=0.0046$ we studied, we no longer observe a FQHS at $\nu=5/2$.
A linear fit to the data passing through the point associated with the $x=0$ reference sample
shows that the gap closes at the extrapolated value of $x \simeq 0.0042$.
We note that the error in $\Delta_{5/2}$ as determined from the Arrhenius fits is estimated to
$\pm 10\%$. However, in Fig.2 there is also scatter in the data possibly caused by small 
variations in the sample densities and slight variations from the
target value of the alloy content $x$.

The aluminum fraction $x$ in our alloy samples is clearly a measure of the added disorder. In 
the literature the most commonly used metric for the disorder is the mobility $\mu$. 
Early work on the $\nu=5/2$ FQHS found that the energy gap of the state correlates well with the
mobility \cite{Pan08}. It was found that a higher $\mu$ resulted in a larger $\Delta_{5/2}$
and the $\nu=5/2$ FQHS does not develop for mobility less than
the threshold value $\mu_C \simeq 10 \times 10^{6}$cm$^2$/Vs.
Later it became apparent that there is in fact a poor correlation between $\Delta_{5/2}$ and $\mu$ \cite{Gamez,Nuebler}. 
Nonetheless, a threshold $\mu_{C}$ below which a $\nu=5/2$ FQHS does not develop
was still identified.
The shaded area of Fig.3 shows the stability region of the $\nu=5/2$ FQHS 
in high quality alloy-free samples at densities $2.65 \times 10^{11} \leq n  \leq 3.2 \times 10^{11}/$cm$^2$
close to that of our samples \cite{Pan11,Gamez,Eisen02,Pan08,Choi,Nuebler,Kumar}. These data are taken from the literature.
A threshold value $\mu_{C} \simeq 7 \times 10^{6}$cm$^2$/Vs for these alloy-free
samples is clearly seen.

Fig.3 also shows that a strong $\nu=5/2$ FQHS with $\Delta_{5/2}=127$~mK develops
in the alloy sample with $\mu=2.2 \times 10^{6}$cm$^2$/Vs. 
This is surprising, since at such a low mobility a $\nu=5/2$ FQHS has never been observed.
Indeed, this mobility is much below the
the previously established $\mu_{C} \simeq 7 \times 10^{6}$cm$^2$/Vs threshold in high quality alloy-free samples. 
We thus found that 
the mobility threshold for a fully quantized $\nu=5/2$ FQHS is significantly lowered in the presence of alloy disorder
and, therefore, the $\nu=5/2$ FQHS is robust to the presence of alloy disorder. 
Furthermore, we conclude that alloy disorder does not appear to be as detrimental to the development of the 
$\nu=5/2$ FQHS as the residual disorder unintentionally added during sample growth. 
The gap $\Delta_{5/2}$ for our alloy samples closes at an extrapolated new threshold of $\mu_C^{alloy} \simeq 1.8 \times 10^{6}$cm$^2$/Vs.

\begin{figure}[b]
 \includegraphics[width=1\columnwidth]{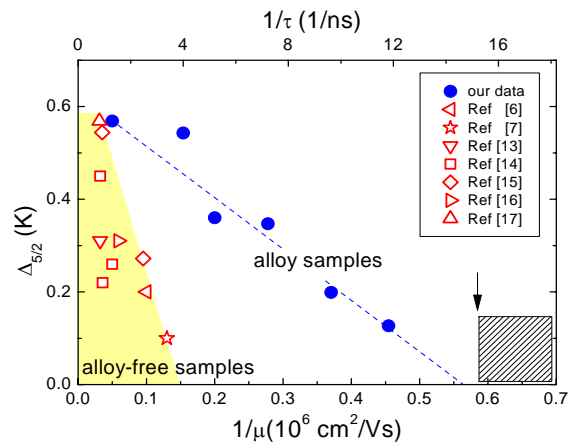}
 \caption{The dependence of $\Delta_{5/2}$ on inverse mobility $1/\mu$ and the electronic scattering
 rate $1/\tau$ of our samples (closed symbols) and of alloy-free samples from the literature with
 densities near $2.9\times10^{11}$/cm$^2$. In our alloy samples the $\nu=5/2$ FQHS survives at
 surprisingly high $1/\mu$ and, therefore, low $\mu$. The arrow indicates a sample in which the
 $\nu=5/2$ FQHS does not develop.
 \label{Fig3}}
 \end{figure}
 
It is important to appreciate that not only our samples have controllably added alloy disorder but, with the exception
of the sample with $x=0.00057$, alloy disorder is the dominant scattering mechanism. This is the case because
the electron scattering rate in our alloy samples $1/\tau$ exceeds the residual scattering rate
of the alloy-free reference sample $1/\tau_{residual}=1.3$ns$^{-1}$ \cite{Gardner}. 
Here $\tau=\mu m/e$ is the transport scattering time.
As seen in Fig.3, $\Delta_{5/2}$ is linearly decreasing with $1/\tau$ with the slope of 41~mK$\cdot$ns.
Furthermore, we find that the gap extrapolates to zero near $1/\tau \simeq 15$~ns$^{-1}$.

Since Al is a neutral impurity, when added to a perfect GaAs crystal, it perturbs the crystal potential on 
sub-nanometer length scale. The alloy disorder we study thus generates a short-range scattering potential \cite{Wanli03}.  
In a recent experiment a different type of short-range disorder, that due to surface roughness scattering, 
was investigated \cite{Pan11}. It was found that in a heterojunction insulated gate field-effect transistor
$\Delta_{5/2}$ increases with a decreasing mobility, a result which is opposite to our findings. 
One reason for this discrepancy is that in Ref.\cite{Pan11} the
electron density is increased as the mobility is decreasing.
The quantitative effect of the disorder on the energy gap in Ref.\cite{Pan11},
therefore, remains difficult to extract. Indeed, the dominance of the short-range scattering
could not be ascertained and the gap altering effects of the combination of changing density and wavefunction confinement
were not disentangled from that of the disorder \cite{Pan11}.
In two other experiments, the effect on the $\nu=5/2$ FQHS of a different type of disorder, 
that due to the remote ionized dopants was investigated \cite{Umansky,Gamez}.
It was found that increasing the level of the remote dopants leads to the strengthening of the $\nu=5/2$ FQHS \cite{Gamez}.
A systematic dependence of the energy gap on overdoping, however, remains unavailable to date.
Our results thus highlight the effect of the short-range alloy disorder on the stability of the $\nu=5/2$ FQHS, 
whereas the effect of other important types of disorder, such as those of the long-range Coulomb potentials of dopants 
and of background impurities, remain unknown.

The lack of correlation of $\Delta_{5/2}$ and $\mu$ in alloy-free samples reported in the literature
remains an outstanding puzzle \cite{Gamez,Nuebler}. 
We propose that such a lack of correlation between $\Delta_{5/2}$ and $\mu$ appears because
a) a well defined $\Delta_{5/2}$ versus $\mu$ correlation exists when only one type of disorder dominates and
a single heterostructure design is employed and
b) for each kind of disorder the $\Delta_{5/2}$ versus $\mu$ functional relationship is different. 
In other words, because high quality alloy-free samples most likely have a different mix of the various disorders
and because $\Delta_{5/2}$ and $\mu$ track differently for each specific type of disorder,
the result is a lack of correlation of $\Delta_{5/2}$ and $\mu$ when an analysis of dissimilar samples is undertaken. 
In contrast, when in a series of similar samples 
one type of disorder dominates, such as in our experiment, $\Delta_{5/2}$ and $\mu$ should be correlated.
We suggest that the quasi-linear correlation of $\Delta_{5/2}$ and $\mu$ in our
alloy samples shown in Fig.3 supports the above hypothesis. This hypothesis, however, remains to be
tested in instances in which various other types of disorder are dominant. A natural consequence
of our analysis is that $\mu$ measured at zero $B$-field, and the single-particle lifetime measured in the
low $B$-field semi-classical regime \cite{coleridge} are poor measures of the impact of the disorder on 
many-body ground states developed at large $B$-fields \cite{Umansky,Pan11,Gamez,Manfra}.
We note that we have measured the single-particle lifetime $\tau_q$ for our series of samples and found 
no obvious correlation with $\Delta_{5/2}$.  The analysis of this data is left to a future publication.

As seen in the inset of Fig.2, above 100~mK there is little or no change with temperature in the magnetoresistance $R_{xx}$ at $\nu=5/2$.
According to the CF description, in this regime the system is described by 
a Fermi sea of the CFs in a zero effective magnetic field \cite{Jain,Halperin}. 
We find the temperature-independent $R_{xx}$ value above 100~mK correlates with the
amount of disorder. The values of $R_{5/2}^{150\text{mK}}$, the saturation value of $R_{xx}$ at 150~mK measured at $\nu=5/2$, 
are listed in the Supplementary Materials section. 
We notice that, $R_{5/2}^{150\text{mK}}$ increases with an increasing $x$. According to the CF
theory, $R_{5/2}^{150\text{mK}}$ is a measure of the scattering of the CF with the impurities \cite{Jain,Halperin,Kang95,JZhang}.
We conclude that the linear increase of $R_{5/2}^{150\text{mK}}$ with $x$ 
is a direct consequence of enhanced scattering rate of the CFs as $x$ increases.
We thus find that at $\nu=5/2$ an increasing alloy disorder has two independent concurring effects: 
it reduces the energy gap of the state and it enhances $R_{5/2}^{150\text{mK}}$, the
$T$-independent $R_{xx}$ at $\nu=5/2$ in the limit of high temperatures. 
 
In an effort to speed up the screening
of the samples and to characterize them at $^3$He refrigerator temperatures at which the $\nu=5/2$ FQHS does
not yet develop, it was proposed that strong $\nu=5/2$ FQHSs develop in samples with low values of the 
$T$-independent $R_{xx}$ measured at $\nu=5/2$ \cite{Manfra}.  
We thus found that such a hypothesis has a natural explanation within the framework of the composite fermion 
theory and, furthermore, that the hypothesis works in samples with alloy disorder.
This hypothesis, however, remains to be further tested in samples with different types of dominating disorder.

We note that recently an alternative method of extracting the energy gap has been proposed \cite{ambrumenil}. 
This model, however, is formulated for the slowly varying potential generated by the remote dopants and
it yet remains to be extended to alloy scattering. There is also effort in understanding short-range scatterers, albeit
so far only for the $\nu=1/3$ FQHS \cite{Sheng}.


In conclusion, we have studied the effect of alloy disorder on the $\nu=5/2$ FQHS in a regime in which 
alloy disorder dominates. 
The gap of the $\nu=5/2$ FQHS closes at unprecedentedly low mobility
which indicates that alloy disorder may not be as detrimental to the formation of the $\nu=5/2$ FQHS
as other types of disorder. Understanding disorder will lead to a better understanding of
other parameters influencing the $\nu=5/2$ FQHS, which will ultimately result in a better engineering of this state.

This work was supported by the DOE BES Experimental Condensed Matter Physics and Synthesis and Processing Science programs
under the award No. DE-SC0006671.

\newpage
\section{Supplementary Material}

We measured a series of quantum well samples with a different aluminum context $x$
in the electron channel. The width of the quantum well is 30~nm in each sample
and certain relevant parameters are listed in Table.I. The capping GaAs layer is 10~nm thick, 
the electron gas is 200~nm deep under the sample surface, and the setback distance
between the dopants and the quantum well is 75~nm on both sides of the well \cite{Gardner9}.
Further details pertaining to the growth procedure and characterization of our samples at 300~mK can be 
found in Ref.\cite{Gardner9}. We note that the densities listed in Table.I may differ 
slightly from those listed in Ref.\cite{Gardner9} because of slight variations due to thermal cycling.

\begin{table}[b]
\caption{A summary of alloy content $x$, electron density $n(10^{11}/\text{cm}^2)$, mobility $\mu(10^6\text{cm}^2/\text{Vs})$, 
scattering rate $1/\tau(1/\text{ns})$, energy gap $\Delta_{5/2}(\text{mK})$ of the $\nu=5/2$ FQHS, 
and $R_{5/2}^{150\text{mK}}(\Omega)$ of the measured samples.}
\begin{ruledtabular}
\begin{tabular}{l c c c c c }
$x$     & $n$  & $\mu$   & $1/\tau$ & $\Delta_{5/2}$ &  $R_{5/2}^{150\text{mK}}$ \\
\hline
0       & 3.08 & 20  & 1.3 & 569  & 27 \\    
0.00057 & 2.91 & 6.5 & 4.0 & 543  & 48 \\
0.00075 & 2.88 & 5.0 & 5.2 & 360  & 83 \\
0.0015  & 2.90 & 3.6 & 7.3 & 347  & 131 \\
0.0026  & 2.70 & 2.7 & 9.7 & 199  & 209 \\
0.0036  & 3.08 & 2.2 & 12  & 127  & 198 \\
0.0046  & 2.82 & 1.7 & 15  & -    & -   \\
\end{tabular}
\end{ruledtabular}
\end{table}

It is instructive to compare the length scales associated with the alloy disorder we study.
The average Al-Al distance within the electron channel of our samples ranges from 5.3-2.7~nm 
when $x$ is between 0.00057-0.0046. In comparison, the unintentionally added charged impurities during the MBE growth
are estimated to have a concentration of about $10^{13}$/cm$^3$, therefore their average separation is close to 0.5~$\mu$m \cite{Umansky9,Manfra9}.

Each sample used in this study is a 4mm$\times$4mm square specimen.
Eight indium-tin contacts are deposited on both the four corners and the middle points of the four edges of each 
specimen and are annealed in an forming gas atmosphere at 400 Celsius degrees for 10 minutes. A silver sinter
heat exchanger is soldered onto each of the ohmic contacts using indium \cite{Nodar119}. The sample and
the silver heat exchangers are housed in a He$^3$ immersion cell.
The role of this cell is to assure electron thermalization \cite{Pan999,Nodar119} and to enable
a magnetic field independent thermometry to these temperatures using a quartz He$^3$ viscometer \cite{Nodar119}.

The preparation of the electronic state is the same for each sample. 
Samples are cooled in our wet dilution refrigerator to about 5~K in the dark.
Samples are then heated up to 10~K and are 
illuminated for 10 minutes with a red light emitting diode (LED).
In order to maintain the same conditions for the LED illumination in different measurements, 
the LED is placed in a similar position with respect to each sample and the same bias current of 1~mA is used.
After illumination, the sample is slowly cooled to 5~K over 2 hours, after which we proceed to condensing the
He$^3$-He$^4$ mixture. Once most of the mixture is condensed, we start filling our He$^3$ immersion cell.
We first obtained magnetotransport data, such as the ones shown in Fig.1 of the main text. A standard
lock-in technique with a low excitation current of 2~nA at 11~Hz was used. The magnetoresistance measurements  
were followed by mobility measurements at zero magnetic field.


\begin{thebibliography}{l}
\bibitem{nextnano} we used the software NextNano3, from the Walter Schottky Institute.
\end{thebibliography}
 \end{document}